\documentclass{article}
\usepackage{amsmath}
\usepackage{amsfonts}
\usepackage{amssymb}

\def\l{\lambda}
\def\p{\partial}

\newtheorem{prop}{Proposition}

\newcommand{\dbar}{\bar{\partial}}

\newcommand{\be}{\begin{equation}}
\newcommand{\ee}{\end{equation}}
\newcommand{\bea}{\begin{eqnarray}}
\newcommand{\eea}{\end{eqnarray}}
\newcommand{\beaa}{\begin{eqnarray*}}
\newcommand{\eeaa}{\end{eqnarray*}}

\newcommand{\nn}{\nonumber}

\begin{document}
\title
{Projective differential geometry of multidimensional 
dispersionless integrable hierarchies}
\author{
L.V. Bogdanov\thanks
{L.D. Landau ITP RAS,
Moscow, Russia}
~and B.G. Konopelchenko
\thanks{Department of Mathematics and Physics ``Ennio De Giorgi", University of
Salento and INFN, Lecce, Italy}
}
\date{}
\maketitle
\begin{center}
\textit{In the memory of S.V. Manakov}
\end{center}
\begin{abstract}
We introduce a general setting for multidimensional dispersionless
integrable hierarchy in terms of differential $m$-form $\Omega_m$
with the coefficients satisfying the Pl\"ucker relations, which is
gauge-invariantly closed and its gauge-invariant coordinates (ratios of coefficients)
are (locally) holomorphic with respect to one of the variables (the spectral variable).
We demonstrate that this form defines a hierarchy
of dispersionless integrable equations
in terms of commuting vector fields locally holomorphic in the spectral variable.
The equations of the hierarchy are  given by the gauge-invariant closedness equations.
\end{abstract}
\section{Introduction}
In this work we develop further the ideas of the work \cite{BK2013}
and introduce a general setting for multidimensional dispersionless
integrable hierarchy in terms of some differential $m$-form $\Omega_m$, in the spirit
of construction of universal Whitham hierarchy given in \cite{Krich94},
which corresponds to the case of $\Omega_2$ with a Hamiltonian reduction.
The coefficients
of this form should satisfy the Pl\"ucker relations (we call it a Pl\"ucker form),
it should be gauge-invariantly closed and its gauge-invariant coordinates (ratios of coefficients)
should be (locally) holomorphic with respect to one of the variables (the spectral variable).
We demonstrate that this form defines a hierarchy
of dispersionless integrable equations
in terms of commuting vector fields locally holomorphic in the spectral variable.
The equations of the hierarchy are  given by the gauge-invariant closedness equations.

First we consider the correspondence between  Pl\"ucker forms and distributions
and demonstrate that involutivity of the distribution is equivalent to the
gauge-invariant closedness equations for Pl\"ucker form.

Then we introduce a spectral variable, suggesting that gauge-independent coordinates
of the form (ratios of coefficients) are holomorphic with respect to one of the variables
and consider some simple examples of integrable systems arising from gauge-invariant
closedness equations.

We demonstrate that nonlinear vector Riemann-Hilbert problem (or $\dbar$-problem)
is a natural tool to construct gauge-invariantly closed Pl\"ucker forms holomorphic
in the complex plane, and show how to obtain polynomial $\Omega_m$ and the corresponding
multidimensional dispersionless hierarchy.
\section{Integrable distributions and closed Pl\"ucker forms}
Let us consider a domain with a set of local coordinates 
$\mathbf{x}=(x_{0},\,x_{1},\,\dots,\,x_{N})$.
Distribution is a $K$-dimensional subspace of the tangent space $\Delta_x\subset T_x$,
depending smoothly on $\mathbf{x}$ (there exists a basis of smooth vector fields).
Involutive distribution is defined by the relation $[\Delta,\Delta]\subset \Delta$, 
where the standard commutator of vector fields is used.
According to
Frobenius theorem, the distribution is integrable (corresponds
to a foliation) $\Leftrightarrow$ the distribution is in involution.


There are several dual formulations of Frobenius theorem in terms of differential forms,
where the subspace in cotangent space dual to the distribution (codistribution) 
is considered.
Here we will consider decomposable differential $m=(N+1-K)$ forms 
(we will call them Pl\"ucker forms)
which are in one-to-one correspondence (up to a gauge) with the codistribution, 
and will formulate the 
property of these forms which is equivalent to the involutivity of original $K$-dimensional 
distribution.

Let us define a Pl\"ucker form as 
an $m$-form
\beaa
\Omega _{m}=\sum_{0\leqslant i_{0}\leq \dots \leqslant i_{m-1}\leqslant N}\pi
_{i_{0}\,i_{1}\,\dots\, i_{m-1}}(x)dx_{i_{0}}\wedge dx_{i_{1}}\wedge
\dots \wedge dx_{i_{m-1}}
\eeaa%
with
coefficients satisfying Pl\"ucker relations (see e.g. \cite{HP})
\beaa
\sum_{l=0}^{m}(-1)^{l}\pi _{i_{0}\,\dots\, i_{m-2}\,j_{l}}
\pi _{j_{0}
\,\dots\,\check j_{l}\,\dots\, j_{m}}=0,
\eeaa%
where indices $i_{p}$ and $j_{p}$ range over all possible values from zero to $N$
and the notation $\check j_{p}$ means the omission of the index. 
Due to Pl\"ucker relations,
the form $\Omega _{m}$ is decomposable
\beaa
\Omega _{m}=\omega _{0}\wedge \dots \wedge \omega _{m-1},
\eeaa
defines a vector subspace in cotangent space and a distribution as a dual object.
It is also easy to construct a Pl\"ucker form for a given distribution, it is defined
up to a gauge.
We will call the Pl\"ucker forms which differ only by a gauge (multiplication by some
function) equivalent.

Having the correspondence between Pl\"ucker forms and distributions, it is natural to
ask a question what property of Pl\"ucker form corresponds to the involutivity of
the distribution. To have a geometrical meaning, this property should
be gauge-invariant.
The answer to this question can be found using the gauge-invariant
closedness conditions for Pl\"ucker forms introduced in \cite{BK2013}.

Let us consider standard closedness equations
\bea
\left[ \frac{\partial \pi _{i_{0}\,i_{1}\,\dots i_{m-1}}}{\partial x_{i_{m}}}%
\right] =0,
\label{closedform}
\eea
where the bracket $[\dots]$ means antisymmetrization
over all indices.
We consider a pair of equations with the choice of indices 
$(0,1,\dots,m-1, q)$,  $(0,1,\dots,m-1, r)$, $q,r\in
\{1,\dots, K=N-m+1\}$,
$q\neq r$,
introducing the basic set of gauge independent affine coordinates 
\beaa
&&
a_{qk}=(-1)^{k}J^{-1}\pi _{0\,\dots\,k-1\,k+1\,\dots\,m-1\,m-1+q},
\\
&&
J=\pi _{0\,1\,...\,m-1} 
\eeaa
where $k\in\{0,\dots,m-1\}$, $q\in\{1,\dots, K=N-m+1\}$. All the affine coordinates
are expressed through the basic set using  
Pl\"ucker relations, and the closedness equations take the form
\bea
\frac{\partial J}{\partial x_{q}}+\sum_{l=0}^{m-1}\frac{\partial (Ja_{ql})%
}{\partial x_{l}}=0,
\quad \frac{\partial J}{\partial x_{r}}+\sum_{l=0}^{m-1}%
\frac{\partial (Ja_{rl})}{\partial x_{l}}=0,
\label{subsystemlin}
\eea
\bea
\frac{\partial a_{qk}}{\partial x_{r}}-\frac{\partial a_{rk}}{\partial
x_{q}}+\sum_{l=0}^{m-1}\left( a_{rl}\frac{\partial a_{qk}}{\partial x_{l}}%
-a_{ql}\frac{\partial a_{rk}}{\partial x_{l}}\right) =0.
\label{subsystem}
\eea
Subsystem (\ref{subsystem}), invariant under the gauge transformations mentioned
above, is the gauge invariant part of the system (\ref{closedform}) for 
Pl\"ucker form \cite{BK2013}; we will call it gauge-invariant closedness equations
for Pl\"ucker form.
Equations (\ref{subsystemlin})
can be viewed as the equations for the gauge variable $J$ which transforms as $%
J\rightarrow \rho J$ under the gauge transformations.  It is a
straightforward check that equations (\ref{subsystemlin}) are compatible due to the
subsystem (\ref{subsystem}). 
We note that the system (\ref{subsystemlin}), (\ref{subsystem}) can be rewritten in the form
\begin{equation}
D_{q}\ln J+\sum_{n=0}^{m-1}\frac{\partial a_{qn}}{\partial x_{n}}=0,\quad
D_{r}\ln J+\sum_{n=0}^{m-1}\frac{\partial a_{rn}}{\partial x_{n}}=0,
\label{Jac}
\end{equation}
\begin{equation}
D_{r}a_{qk}-D_{q}a_{rk}=0,\quad k=0,\dots,m-1
\end{equation}%
where $D_{q}$ and $D_{r}$, $q,r\in
\{1,\dots, K\}$,  are vector fields
\begin{equation}
D_{q}=\frac{\partial }{\partial x_{m}}+\sum_{n=0}^{m-1}a_{qn}\frac{%
\partial }{\partial x_{n}},\quad D_{r}=\frac{\partial }{\partial x_{m+1}}%
+\sum_{n=0}^{m-1}a_{rn}\frac{\partial }{\partial x_{n}}.
\label{vector_fields}
\end{equation}
Equations of the type (\ref{Jac}) 
were considered in \cite{LVB10MS}, \cite{LVB11} as equations
for the Jacobian of solutions of linear equations defined by vector fields.
So the subsystem (\ref{subsystem}) 
is equivalent to the commutativity $\left[ D_{q},D_{r}%
\right] =0$ of vector fields $D_{q}$ and $D_{r}.$

Considering the set of gauge-invariant closedness equations
(\ref{subsystem}), we come to the following
conclusions:
\begin{prop}
The set of gauge-invariant closedness equations (\ref{subsystem})
is equivalent to the existence of a gauge, in 
which the Pl\"ucker form is closed in the standard sense. 
\end{prop}
The gauge variable $J$ corresponding to closed form
is defined by linear equations (\ref{subsystemlin}).
\begin{prop} The 
Pl\"ucker form is gauge-invariantly closed $\Leftrightarrow$ the corresponding distribution is in involution
\label{involution}
\end{prop}
The
Proposition \ref{involution} can be easily proved by purely geometrical means, using the Frobenius theorem
and some well-known properties of differential forms.
It is known that closed decomposable differential form has a decomposition in terms
of exact 1-forms (see e.g. \cite{St,Ram}). 
The fact that the Pl\"ucker form is decomposable implies that
(gauge invariantly) closed Pl\"ucker form possesses a decomposition into exact 1-forms
(up to a gauge).
\begin{prop}
Gauge-invariantly closed Pl\"ucker form can be represented as
\bea
\Omega_{m}=g df_{0}\wedge \dots \wedge df_{m-1},
\label{decomposition}
\eea
where $g$, $f_0,\dots,f_{m-1}$ are some functions.  
\end{prop}
\subsection*{Examples of closedness equations}
Let us consider some simple examples of closedness equations
(\ref{subsystemlin}), (\ref{subsystem})
for Pl\"ucker forms,
for more detail see \cite{BK2013}. We will use the representation
\begin{equation}
\Omega _{m}=J\tilde{\Omega }_{m}=J(dx_0\wedge \dots \wedge dx_{m-1} + \dots),
\label{tildeform}
\end{equation}%
where $J=\pi _{0\,1\,...\,m-1}$ is a gauge-dependent variable, and the Pl\"ucker form 
$\tilde{\Omega }_{m}$ is gauge-independent, it may be considered as the initial form in affine gauge.

The simplest case corresponds to $m=1$, let us take $N=2$:
\beaa
\Omega_1=J(dx_0-a_{10}dx_1-a_{20}dx_2).
\eeaa
In this case we have no Pl\"ucker relations, and the closedness equations are
\bea
&&
\frac{\partial J}{\partial x_{1}}+\frac{\partial (Ja_{10})}{\partial x_{0}}%
=0,\quad \frac{\partial J}{\partial x_{2}}+\frac{\partial (Ja_{20})}{%
\partial x_{0}}=0,
\\
&&
\frac{\partial a_{10}}{\partial x_{2}}-\frac{%
\partial a_{20}}{\partial x_{1}}+a_{20}\frac{\partial a_{10}}{\partial x_{0}}%
-a_{10}\frac{\partial a_{20}}{\partial x_{0}}=0.
\label{subN2}
\eea
For the case $m=2$, 
$N=3$ we have
\bea
\Omega _{2} &=&J(dx_{0}\wedge dx_{1}-a_{11}dx_{0}\wedge
dx_{2}-a_{21}dx_{0}\wedge dx_{3}+a_{10}dx_{1}\wedge dx_{2}+ \\
&&+a_{20}dx_{1}\wedge dx_{3}-(a_{11}a_{20}-a_{10}a_{21})dx_{2}\wedge dx_{3}),
\label{form2}
\eea
where the Pl\"ucker relations are taken into account in the last term.
The closedness equations read
\bea
\frac{\partial J}{\partial x_{2}}+\sum_{m=0}^{1}\frac{\partial (Ja_{1m})}{%
\partial x_{m}}=0,\quad \frac{\partial J}{\partial x_{3}}+\sum_{m=0}^{1}%
\frac{\partial (Ja_{2m})}{\partial x_{m}}=0,
\label{subN3a}
\eea
\bea
\bigskip \frac{\partial a_{1k}}{\partial x_{3}}-\frac{\partial a_{2k}}{%
\partial x_{2}}+\sum_{l=0}^{1}\left( a_{2l}\frac{\partial a_{1k}}{\partial
x_{l}}-a_{1l}\frac{\partial a_{2k}}{\partial x_{l}}\right) =0,\qquad k=0,1.
\label{subN3}
\eea

\section{The spectral variable. Dispersionless integrable systems}
To introduce dispersionless integrable systems, we need 
the solutions of gauge-invariant closedness equations (\ref{subsystem})
holomorphic with respect to one variable (the spectral variable).
Let us consider  (gauge invariantly)
closed Pl\"ucker form $\Omega_m$
with affine coordinates (ratios of coefficients) holomorphic 
with respect to $\lambda=x_0$ in some complex domain. This form defines a hierarchy
of dispersionless integrable equations
in terms of commuting vector fields locally holomorphic in $\lambda$.
The equations of the hierarchy are  given by the gauge-invariant closedness equations.

More specifically, we consider the forms meromorphic in the complex plane
(in the affine gauge).

This setting for $m=2$ can be reduced to Whitham hierarchy (Krichever \cite{Krich94}), 
for $m=3$ to heavenly equation hierarchy (Takasaki \cite{Takasaki89}, \cite{Takasaki89a}) and 
connected Dunajski
equation hierarchy \cite{Dun02}, \cite{BDM07}. 

Important reductions are volume (or area)
conservation corresponding to closedness in the affine gauge
($J=1$) and hyper CR (Cauchy-Riemann) reduction $\Omega_m\wedge d\lambda=0$.

Most known examples correspond to the case when there exists polynomial
(or Laurent polynomial) set of affine coordinates (affine gauge).
Below we will restrict ourselves to the polynomial case. Examples corresponding
to Laurent polynomials were considered in \cite{BK2013}, \cite{LVB10Toda}.

A closely related geometric picture of dispersionless integrable systems
in terms of coisotropic deformations was introduced in \cite{KF}, \cite{KO}, and
connection between this picture and the setting of the present work was discussed
in \cite{BK2013}.
\subsubsection*{Polynomial $\Omega_1$}
The case $m=1$ is non-generic, in this case there is no hierarchy of commuting systems, 
and it is not clear how to
solve it in general. The  equations corresponding to this case were considered in the framework
of inverse scattering method with variable spectral parameter \cite{BZM87}. 
The gauge-invariant closedness condition for the form
\beaa
\Omega_1=J(d\lambda+udx-
(1+v\lambda+\lambda ^{2})dy),
\eeaa
which is given by equation (\ref{subN2}), 
leads to the Liouville equation 
\be
\varphi_{xy}=e^\varphi,
\label{Liouville}
\ee
where $u=\tfrac{1}{2}e^\varphi$, the Lax pair for this equation reads
\beaa
&&
\partial_x\psi=u\partial_\lambda\psi
\\
&&
\partial_y\psi=-(1+v\lambda+\lambda^2)\partial_\lambda\psi.
\eeaa
The case of third order polynomial
\beaa
\Omega_1=J(d\lambda+udx-
(1+w\lambda+w'\lambda ^{2}+\lambda ^{3})dz)
\eeaa
leads to `higher Liouville
equation' introduced in \cite{BZM87},
\be
\varphi_{xxz}-\varphi_{xz}\varphi_x=\tfrac{3}{2}e^\varphi,
\label{HL}
\ee
where $w=\varphi_z$, 
$w'=e^{-\varphi}\varphi_{xz}$, the Lax pair for this equation
is
\beaa
&&
\partial_x\psi=u\partial_\lambda\psi
\\
&&
\partial_z\psi=-(1+w\lambda+w'\lambda^2+\lambda^3)\partial_\lambda\psi.
\eeaa
We should emphasize that equations (\ref{Liouville}), (\ref{HL}) are not 
commuting and they do not belong to a hierarchy. It is also possible to introduce
`higher Liouville equations' of arbitrary order. 
\subsubsection*{Polynomial $\Omega_2$}
For the case $m=2$, 
$N=3$ we consider the form (\ref{form2}) with polynomial coefficients
\beaa
&&
a_{10}=u_{0}(x),\quad a_{11}=u_{1}(x)+\lambda , 
\nn\\
&&
a_{20}=v_{0}(x)+\lambda v_{1}(x),\quad
a_{21}=v_{2}(x)+\lambda v_{3}(x)+\lambda ^{2}.
\eeaa
Considering gauge-invariant closedness equations (\ref{subN3}) and denoting 
$x=x_{1}$, $y=x_{2}$, $t=x_{3}$ (see \cite{BK2013} for more detail), 
one gets the Manakov-Santini system \cite{MS06}, \cite{MS07}
\bea
u_{xt}+u_{yy}+(uu_{x})_x+v_{x}u_{xy}-v_{y}u_{xx}=0,
\nn
\\
v_{xt}+v_{yy}+uv_{xx}+
v_{x}v_{xy}-v_{y}v_{xx}=0.
\label{MS}
\eea
Considering more variables and higher order polynomials, it is possible
to introduce the form $\Omega_2$ corresponding to the Manakov-Santini
hierarchy \cite{LVB09}.
\subsubsection*{Reductions}
There are two important classes of reductions of dispersionless systems
(hierarchies)
defined in terms of the
form $\Omega _{m}$.
\begin{enumerate}

\item The form $\Omega _{m}$ is closed in standard sense in affine gauge ($J$=1). In general, the closedness in the affine gauge
leads to representation of the hierarchy 
in terms of volume-preserving (divergence-free) vector fields.
In this case the equations of gauge-invariant closedness (\ref{subsystem}) are complemented by equations, implied by linear subsystem (\ref{subsystemlin})
for $J=1$, 
\beaa
\sum_{l=0}^{m-1}\frac{\partial a_{ql}%
}{\partial x_{l}}=0, \quad q\in\{1,\dots,K\}.
\eeaa
For the Manakov-Santini system this reduction leads to the condition
$v=0$, defining the dKP equation.

\item Reduction $\Omega _{m}\wedge d\lambda=0$. In this case it is possible to
consider $\Omega_m=d\lambda\wedge\Omega'_{m-1}$ with $\Omega'_{m-1}$
not containing $d\lambda$ and gauge-invariantly closed, and $\lambda$
plays a role of a parameter. 
Vector fields do not contain
a derivative over spectral variable. 
In the case of Manakov-Santini system (\ref{MS}) this reduction
leeds to the condition $u=0$ and the equation
\beaa
v_{xt}+v_{yy}+
v_{x}v_{xy}-v_{y}v_{xx}=0
\eeaa
considered in \cite{MASh02}, \cite{Pavlov03}, \cite{MASh04}.
For general $\Omega _{m}$ this reduction defines the hyper CR (Cauchy-Riemann) hierarchies.
\end{enumerate}


\section{Nonlinear Riemann-Hilbert problem}

Let us consider a form
\beaa
\Omega _{m}=d\Psi ^{0}\wedge d\Psi ^{1}\wedge \dots d\Psi^{m-1}
=J\tilde\Omega_m,
\eeaa
where  $\Psi ^{k}$ are
some functions  (series in $\lambda$),  $J$ is some coefficient of the form in coordinates $\mathbf{x}$,
$\tilde\Omega_m$ is a gauge-invariant (affine) factor.  It is easy to see that 
$\Omega _{m}$ is a closed Pl\"ucker form, and the only thing we need 
to construct a solution of some dispersionless integrable system
is to provide definite analytic properties of the affine factor.

\textbf{Question}
{\em How to provide some simple analytic properties
of the affine factor? What kind of functions $\Psi ^{k}$ correspond
to a polynomial affine factor?}


It is 
easy to see that $\tilde\Omega_m$ is invariant under diffeomorphism
\beaa
(\Psi ^{0},\Psi ^{1},\dots, \Psi^{m-1})\rightarrow 
\mathbf{F}(\Psi ^{0},\Psi ^{1},\dots, \Psi^{m-1})
\eeaa
Let the functions 
$\Psi ^{k}$ be holomorphic (meromorphic) inside and outside the unit circle (or some
curve in the complex plane),
having a discontinuity on it. If they satisfy a nonlinear vector
Riemann-Hilbert problem on the unit circle
\bea
(\Psi ^{0},\Psi ^{1},\dots, \Psi^{m-1})_\text{in} =
\mathbf{F}(\Psi ^{0},\Psi ^{1},\dots, \Psi^{m-1})_\text{out},
\label{RH}
\eea
then the affine factor $\tilde\Omega_m$ is holomorhic (meromorphic).

Thus nonlinear vector
Riemann-Hilbert problem  gives a tool to construct closed
Pl\"ucker forms with holomorphic (meromorphic) affine factor, generating
commuting vector fields with holomorphic (meromorhic) coefficients.

Equivalently, it is possible to use a nonlinear vector $\dbar$-problem 
in some domain of the complex plane
\beaa
\dbar \Psi ^{k}={F}^k(\Psi ^{0},\Psi ^{1},\dots, \Psi^{m-1}),
\quad 0\leqslant k \leqslant m-1.
\eeaa
It provides the analiticity of affine factor $\tilde\Omega_m$
in the domain.
\subsection*{General hierarchy for the polynomial case}
We will demonstrate how for a special choice of structure of functions
$\Psi ^{k}$, using nonlocal Riemann-Hilbert problem,  it is possible
to obtain polynomial affine factor and the corresponding hierarchy.
We consider the formal series
\bea
&&
\Psi^0=\lambda+\sum_{n=1}^\infty \Psi^0_n(\mathbf{t}^1,\dots,\mathbf{t}^{m-1})\l^{-n},
\label{form0}
\\&&
\Psi^k=\sum_{n=0}^\infty t^k_n (\Psi^0)^{n}+
\sum_{n=1}^\infty \Psi^k_n(\mathbf{t}^1,\dots,\mathbf{t}^{m-1})(\Psi^0)^{-n},
\label{formk}
\eea
where $1\leqslant k\leqslant m-1$, depending on 
$m-1$ infinite sequences of independent variables
$\mathbf{t}^k=(t^k_0,\dots,t^k_n,\dots)$, $t^k_0=x_k$, $\lambda=x_0$.

Let us consider the Riemann-Hilbert problem (\ref{RH}) on the unit circle,
where the functions $\Psi ^{k}$  are analytic inside and outside the circle and 
in the neighborhood of infinity
are given by the series of the form
(\ref{form0}), (\ref{formk}), 
Then the form $\tilde\Omega_m$ is analytic in the complex plane, morover, due to the structure
of the series it is polynomial. Thus it corresponds to dispersionless hierarchy  for
the polynomial case.
Equations of the hierarchy are generated by the relation (see \cite{BDM07}, \cite{LVB09})
\beaa
\left( J^{-1}d\Psi ^{0}\wedge d\Psi ^{1}\wedge \dots d\Psi
^{m-1}\right) _{-}=0
\eeaa
where $\left( {\cdots}\right) _{-}$ denotes the projection on the part of 
$\left({\cdots}\right)$ 
with negative powers in $\lambda$ and
$
J=\pi _{0\,1\,\dots\,m-1}=\det
(\p_{l}\Psi^{k})_{k,l=0,\dots,m-1}.
$
Generating relation represents analyticity condition for affine factor
of the closed
Pl\"ucker form, which can be provided by the Riemann-Hilbert problem (\ref{RH}).


Using the Jacobian matrix
$$
(\text{Jac}_0)=\left(\frac{D(\Psi^0,\dots,\Psi^{m-1})}
{D{({x_0,\dots,x_{m-1}})}}\right),\quad \det(\text{Jac}_0)=J,
$$
it is possible to write Lax-Sato equation of the hierarchy in the form
\bea
&&
\partial^k_n\mathbf{\Psi}=\sum_{i=0}^{m-1}
\left((\text{Jac}_0)^{-1})_{ik} (\Psi^0)^n)\right)_+
{\partial_i}\mathbf{\Psi},\quad
1\leqslant k \leqslant m-1,
\label{genSato}
\eea
where $1\leqslant n < \infty$,
$\mathbf{\Psi}=(\Psi^0,\dots,\Psi^{m-1})$. 
First flows of the hierarchy read
\bea
\partial^k_1\mathbf{\Psi}=(\lambda \partial_k-\sum_{p=1}^{m-1} 
(\partial_k u_p)\partial_p-
(\partial_k u_0)\partial_\lambda)\mathbf{\Psi},\quad 1\leqslant k\leqslant m-1,
\label{genlinear}
\eea
where $u_0=\Psi^0_1$,
$u_k=\Psi^k_1$, $1\leqslant k\leqslant m-1$.


A compatibility condition for any pair of linear equations  
(e.g., with $\partial^k_1$ and $\partial^q_1$, $k\neq q$)
implies closed nonlinear 
N-dimensional  system of PDEs for the set of functions $u_k$, $u_0$,
which can be written in the form
\bea
&&
\partial^k_1\p_q\hat u-\partial^q_1\p_k\hat u+[\p_k \hat u,\p_q \hat u]=
(\p_k u_0)\p_q-(\p_q u_0)\p_k,
\nn\\
&&
\partial^k_1\p_q u_0 - \partial^q_1\p_k u_0 + (\p_k \hat u)\p_q u_0 -
(\p_q \hat u)\p_k u_0=0,
\label{Gensystem}
\eea
where $\hat u$ is a vector field, $\hat u=\sum_{p=1}^{m-1} u_p \p_p$. 
For $m=3$ this system after volume-preservation reduction corresponds 
to the Dunajski system (generalizing heavenly equation).

\section*{Acknowledgements}
The research of 
LVB was partially supported  by the President of Russia grant 6170.2012.2 (scientific schools),
the research of BGK was partially supported by the PRIN 2010/2011 grant 2010JJ4KBA.003.

\end{document}